\documentclass[prd,preprint,nofootinbib]{revtex4}
\usepackage[latin9]{inputenc}
\usepackage{amsmath}
\usepackage{amssymb}
\usepackage{graphicx}
\usepackage{color}

\makeatletter
\@ifundefined{textcolor}{}
{%
 \definecolor{BLACK}{gray}{0}
 \definecolor{WHITE}{gray}{1}
 \definecolor{RED}{rgb}{1,0,0}
 \definecolor{GREEN}{rgb}{0,1,0}
 \definecolor{BLUE}{rgb}{0,0,1}
 \definecolor{CYAN}{cmyk}{1,0,0,0}
 \definecolor{MAGENTA}{cmyk}{0,1,0,0}
 \definecolor{YELLOW}{cmyk}{0,0,1,0}
}

\makeatother

\begin{document}

\title{Big bang nucleosynthesis constraints on varying electron mass solution to the Hubble tension}

\author{Osamu Seto}
\email{seto@particle.sci.hokudai.ac.jp}
\affiliation{Department of Physics, Hokkaido University, Sapporo 060-0810, Japan}

\author{Yo Toda}
\email{y-toda@particle.sci.hokudai.ac.jp}
\affiliation{Department of Physics, Hokkaido University, Sapporo 060-0810, Japan}

\begin{abstract}
A cosmological model with a time-varying mass of electrons seems a promising solution for the so-called Hubble tension. We examine the big bang nucleosynthesis (BBN) constraints on the time-varying electron mass model, because a larger electron mass gives rise to the smaller weak interaction rate for the proton and neutron conversion, which could affect the light element abundance. Additionally, different inferred cosmological parameters, primarily baryon asymmetry, keeping the cosmic microwave background power spectrum unchanged could affect the abundance of light element. We find that the resultant proton-to-neutron ratio is not so much sensitive with respect to the electron mass, because the change of weak interaction rate becomes important after the cosmic temperature becomes lower than the electron mass,
while the slightly smaller present Hubble parameter and the electron mass are indicated if the BBN data are taken into account. We also find that the baryon density is more stringently constrained by the baryon acoustic oscillation data rather than the BBN. We have derived the ratio of the electron mass at early Universe and the present $m_{e}/m_{e0} = 1.0028 \pm 0.0064$  and $H_0 = 68.0\pm 1.1 \mathrm{km/s/Mpc}$. 
\end{abstract}
\preprint{EPHOU-22-011}

\maketitle





\section{Introduction}


The cosmic expansion is one of the most fundamental properties of our Universe.
The present expansion rate called the Hubble constant $H_0$ has been measured by various methods.
The direct measurements of $H_0$ with low redshifts distant ladders report larger values of $H_0$ than the cosmological estimation $H_0 =67.4\pm 0.5$ km/s/Mpc with the temperature and polarization anisotropy of the cosmic microwave background (CMB) by Planck (2018)~\cite{Aghanim:2018eyx}.
The SH0ES collaboration reported $H_0 = 73.04\pm 1.04 $ km/s/Mpc
 by using Cepheids and type Ia supernovae as the standard candle
 in Ref.~\cite{Riess:2021jrx}.
Another local measurement using the Tip of the Red Giant Branch 
 as distance ladders~\cite{Freedman:2019jwv} and 
the H0LiCOW collaboration by the measurements of lensed quasars~\cite{Wong:2019kwg} also report a larger value of $H_0$ than the Planck results.
The discrepancy between various low-redshift measurements of the Hubble constant and
 the value inferred from the CMB temperature anisotropy by Planck (2018) is now regarded as 
 the Hubble tension or the $H_0$ tension.

This problem has been the object of many attempts~\cite{DiValentino:2021izs,Abdalla:2022yfr}.
One of the simplest approaches is to introduce additional relativistic
degrees of freedom parametrized by $\Delta N_\mathrm{eff}=N_\mathrm{eff}-N_\mathrm{eff}^\mathrm{SM}$,
 where $N_\mathrm{eff}$ is the total effective relativistic
degrees of freedom and $N_\mathrm{eff}^\mathrm{SM}$ is its predicted value by the standard model (SM) of particle physics~\cite{Mangano:2005cc,Escudero:2020dfa,Akita:2020szl,Bennett:2020zkv}.
The measured angular size $\theta_{*}\equiv r_{s*}/D_{M*}$ of the acoustic scale of the CMB
infers a shorter angular diameter distance $D_{M*} \propto 1/H_0$, if the sound horizon scale $r_{s*}$ at the recombination epoch is shorter by the extra energy density with a larger $N_{\mathrm{eff}}$.
The preferred value has been suggested as $0.2 \lesssim \Delta N_\mathrm{eff} \lesssim 0.5$ based on the CMB, the baryon acoustic oscillation (BAO) and the SH0ES in Ref.~\cite{Aghanim:2018eyx}. 
If we also account for the successful Big bang nucleosynthesis (BBN) in the analysis,
a smaller value as $0.2 \lesssim \Delta N_{\mathrm{eff}} \lesssim 0.4$ is preferred~\cite{Seto:2021xua} because the extra energy density could speed up the cosmic expansion at the BBN epoch and alter the resultant Helium mass fraction $Y_P$.
Such additional energy density is limited not only by the BBN, but by the CMB itself as well.
If the cosmic expansion rate at the recombination time is increased by extra energy densities to reduce the sound horizon scale, then simultaneously the relative scale between the sound horizon and the photon diffusion (Silk damping) length is also changed. As a result, all models of this class are confronted with the limitation~\cite{Knox:2019rjx}.

A cosmological model with a time-varying electron mass $m_e$~\cite{Barrow:2005qf,Barrow:2005sv} can effectively reduce the sound horizon scale~\cite{Planck:2014ylh,Hart:2019dxi,Sekiguchi:2020teg,Solomon:2022qqf}.
As the energy level of hydrogen $E^\mathrm{H} \propto m_e$ and the Thomson scattering cross section $\sigma_T \propto 1/m_e^2$, the time-varying electron mass could reduce both the sound horizon scale and the Silk damping scale, which does not affect the power spectrum of the CMB temperature anisotropy.

In this paper, we examine other possible consequences of the time-varying electron mass model in which the mass of the electron has been about a few percent larger than the present value before the recombination. Such one important event in the early Universe is the BBN where light elements were synthesized. 
The helium mass fraction has been reported as $Y_p = 0.2449 \pm 0.0040$ from the recent observation data~\cite{Aver:2015iza}. 
The deuterium abundance $D/H$ has been measured as $(D/H)  = (2.527 \pm 0.030)\times 10^{-5}$~\cite{Cooke:2017cwo}.
However, if the electron mass is larger than the present value, then the abundance of synthesized light elements would be altered~\cite{Yoo:2002vw,Ichikawa:2006nm,Scoccola:2008yf,Landau:2008re,Hoshiya:2022mts}. One reason is that the decay rate of the neutron $n$ to the proton $p^+$, electron $e^-$ and neutrino $\nu$ also depends on the electron mass as~\cite{Dicus:1982bz}
\begin{align}
\Gamma(n \rightarrow p^+ e^- \nu) &= \frac{G_F^2}{2\pi^3}(1+3g_A^2)m_e^5 \lambda_0(q) ,
\label{eq:Gamma_n} \\
\lambda_0(q) &= \int_1^q dx x (x-q)^2 (x^2 -1 )^{1/2}, 
\end{align}
 with $q = Q/m_e$ 
 where $G_F$ is the Fermi constant, $g_A$ is the axial-vector coupling of the nucleon and
 $Q=m_n-m_p$ is the difference between the proton mass $m_p$ and the neutron mass $m_n$. 
The rate of weak interactions such as $\nu + n \leftrightarrow p^+ + e^-$ also depends on the electron mass~\cite{Weinberg:1972kfs}. 
Thus, it causes earlier (later) decoupling of $n \leftrightarrow p$ interactions for a larger (smaller) electron mass and results the change of the neutron to proton ratio.
The prediction of the BBN is sensitive to the variation of the neutron decay rate and the weak interaction rate~\cite{Iocco:2008va,Cyburt:2015mya}.
We derive the constraints on the variation of the electron mass in the early Universe, assuming that 
the electron mass at the BBN time is the same as at the recombination epoch and that the other fundamental parameters such as the fine structure constant and the Fermi constant are not varied.


\section{Data and Analysis}

\label{sec:analysis}

We conduct a Markov-chain Monte Carlo analysis on the time-varying
electron mass model. We use the public Markov-chain Monte Carlo code \texttt{CosmoMC-planck2018}~\cite{Lewis:2002ah}.
The change of the electron mass is implemented in the neutron decay rate
in the BBN era as 
\begin{equation}
\Gamma(n\rightarrow p^{+}e^{-}\nu_{e})_{\mathrm{BBN}}=\Gamma_{n0}\left(\frac{m_{e}}{m_{e0}}\right)^{5}\frac{\lambda_{0}(Q/m_{e})}{\lambda_{0}(Q/m_{e0})},
\end{equation}
through Eq.~(\ref{eq:Gamma_n}) with the present values of the neutron
lifetime $\tau_{n0}=1/\Gamma_{n0}=879.4$ second as well as in the $n\rightarrow p^+$ weak interaction rate through its phase space integral~\cite{Weinberg:1972kfs}
\begin{equation}
\lambda(n\rightarrow p^+) \propto \left(
\int_{-\infty}^{-m_e-Q}\frac{x^2 (Q+x)^2 \sqrt{1-\frac{m_e^2}{(Q+x)^2}}}{\left(e^{\frac{x}{T_{\nu}}}+1\right) \left(e^{-\frac{Q+x}{T}}+1\right)}dx
+\int_{m_e-Q}^{\infty}\frac{x^2 (Q+x)^2 \sqrt{1-\frac{m_e^2}{(Q+x)^2}}}{\left(e^{\frac{x}{T_{\nu}}}+1\right) \left(e^{-\frac{Q+x}{T}}+1\right)}dx \right),
\label{Eq:phaseint}
\end{equation}
 with $T_{\nu}$ being the temperature of neutrinos, 
and similar in the $p^+\rightarrow n$ weak interaction rate.
From now on, $m_{e}$ denotes the electron mass in the early Universe and we use the present electron mass $m_{e0}=511$ keV~\cite{Workman:2022zbs}.


We analyze the models by referring to the following cosmological observation
datasets. We include both temperature and polarization likelihoods
for high $l$ \texttt{plik} ($l=30$ to $2508$ in TT and $l=30$
to $1997$ in EE and TE) and low$l$ \texttt{Commander} and lowE \texttt{SimAll}
($l=2$ to $29$) of Planck (2018) measurement of the CMB temperature
anisotropy~\cite{Aghanim:2018eyx}. We also include Planck lensing~\cite{Aghanim:2018oex}
and data of the BAO from 6dF~\cite{Beutler:2011hx}, DR7~\cite{Ross:2014qpa},
and DR12~\cite{Alam:2016hwk}. We use the datasets of the $Y_{P}$~\cite{Aver:2015iza}
and $D/H$ measurements~\cite{Cooke:2017cwo} to impose the BBN constraints.

\section{Result and discussion}

\label{sec:result}

\begin{figure}[htbp]
\centering
\includegraphics[width=14.5cm]{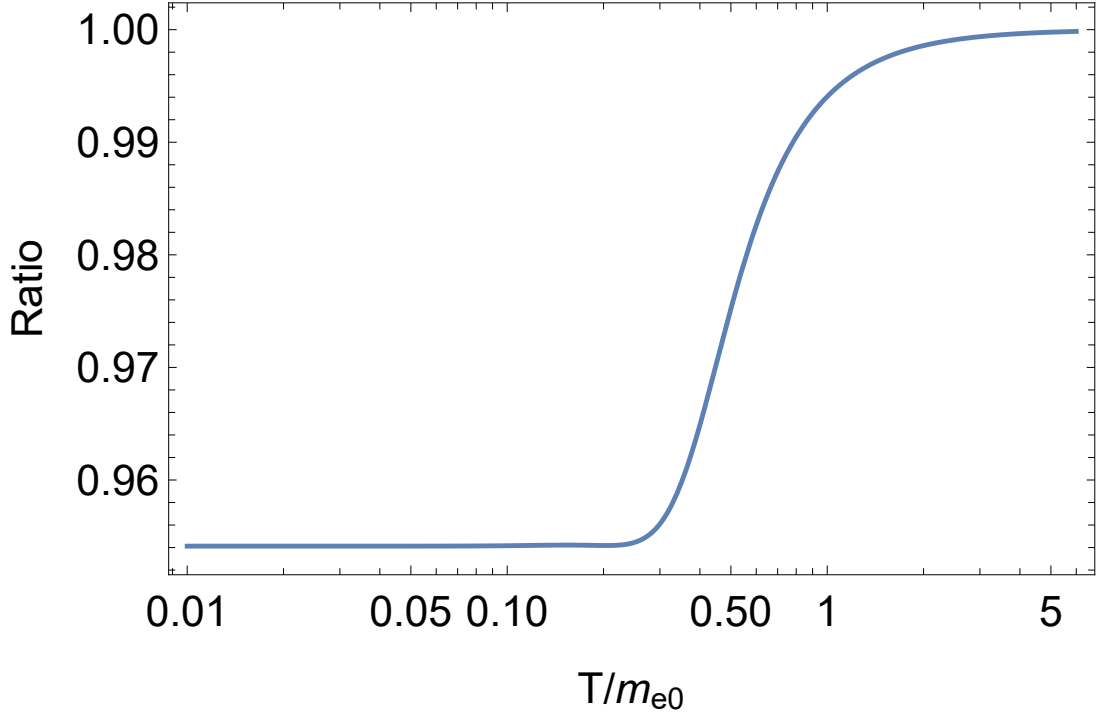} 
\caption{The suppression of $\lambda(n\rightarrow p^+)$ for $m_e=1.03 \times m_{e0}$ as a function of $T/m_{e0}$. }
\label{Fig:wratio} 
\end{figure}

We calculate light elements abundance by using \texttt{PArthENoPE3.0-Standard}~\cite{Gariazzo:2021iiu} with the above modification.
The change of phase space in Eq.~(\ref{Eq:phaseint}) by varying the electron mass induces the enhancement or suppression of the $p^+ \leftrightarrow n$ interaction rate.
We show that the value of $\lambda(n\rightarrow p^+)$ for $m_e= 1.03 \times m_{e0}$ is normalized by that for $m_{e0}$ in Fig.~\ref{Fig:wratio}.
As can be seen in Fig.~\ref{Fig:wratio},
 the interaction rate is insensitive with respect to $m_e$ at a high temperature $T \gg m_e \sim m_{e0}$,
 while the interaction rate changes at a low temperature $T \ll m_e \sim m_{e0}$
 where it is dominated by the beta decay of neutron $n \rightarrow p^+ e^- \bar{\nu}$ with the decay rate given
 by Eq.~(\ref{eq:Gamma_n}).
Indeed, for the case shown in Fig.~\ref{Fig:wratio}, the neutron decay rate is reduced by about $4.6 \%$ due to the increasing of the electron mass $m_e = 1.03 \times m_{e0}$.
We notice that the effect on the $p^+ \leftrightarrow n$ reaction rate by different electron mass
 appears only after 
\begin{equation}
T \lesssim m_{e0}.
\label{Eq:Tlessme}
\end{equation}
The neutron-to-proton ratio
\begin{equation}
\frac{n}{p} \simeq \exp\left(-\frac{Q}{T_D}\right) \simeq \frac{1}{6},
\end{equation}
is realized with the decoupling temperature
 of $p^{+} \longleftrightarrow n$ interaction, $T_{D}\simeq 0.8$ MeV~\cite{Kolb:1990vq}.
By comparing with Eq.~(\ref{Eq:Tlessme}), we find that the effect of varying electron mass
 just starts to appear and is not yet so significant when the $p^{+} \longleftrightarrow n$ conversion freezes out at $T\simeq T_D$.

So far, we have found that the resultant proton-to-neutron ratio is not affected much
 for a fixed baryon-to-photon ratio $\eta$.
The time-varying electron mass model looks an effective solution to
the Hubble tension because the electron mass variation effect on
the CMB power spectrum can be absorbed by varying the inferred baryon
density parameter $\Omega_{b}h^{2}$ and the matter density parameter
$\Omega_{m}h^{2}$ due to the parameter degeneracy~\cite{Sekiguchi:2020teg}. 
The other way to affect light elements abundance by the variation of the electron mass is
 through the change of the inferred baryon asymmetry $\Omega_{b}h^{2}$.
As is well known, the $D/H$ would reduce as the inferred baryon asymmetry $\Omega_{b}h^{2}$ becomes larger.

\begin{figure}[thbp]
\centering
\includegraphics[width=\hsize]{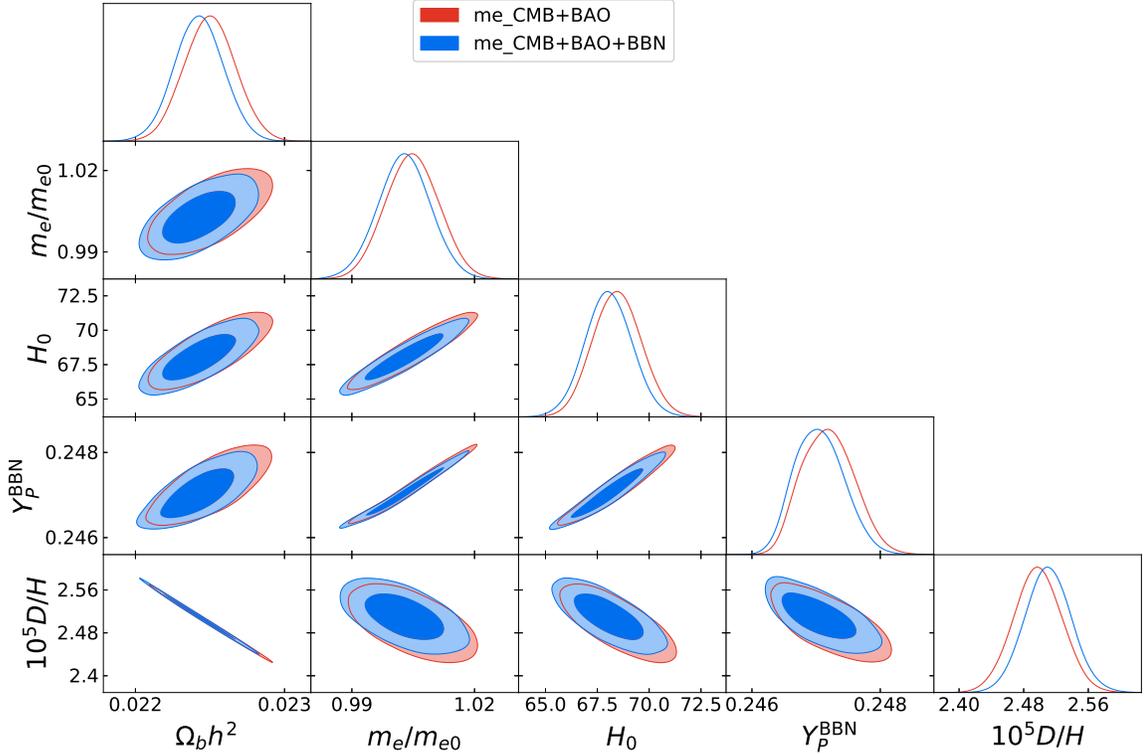} 
\caption{Posterior distributions of light elements abundance $Y_{P}$ and $D/H$, $m_{e}/m_{e0}$,
and $H_{0}$ for several values of $m_{e}/m_{e0}$. }
\label{Fig:MC1} 
\end{figure}

This can be seen in the 2D posterior distributions of cosmological and BBN parameters of 
Fig.~\ref{Fig:MC1}.
We do not find the significant difference for varying electron mass model by 
including BBN data because the $p/n$ ratio and the resultant helium fraction are 
not affect much by the variation of the electron mass, as has been discussed. 
Since the $\Omega_{b}h^{2}$ in time-varying electron mass models has been adjusted 
to fit the CMB spectrum, for the same inferred somewhat smaller $\Omega_{b}h^{2}$,
a larger $D/H$ is indicated.

To see the detail of $\chi^{2}$, the predicted $H_{0}$, $\Omega_{m}$, $Y_{P}$ and
$D/H$ of the best fit point in the varying $m_{e}$ model and the $\Lambda$CDM model 
are listed in Table~\ref{Table:chi^2} for the dataset of CMB, BAO and BBN.
We find, between the constraints on the varying $m_e$ model, that the BAO constraint is more stringent 
than the BBN.
\begin{table}[htbp]
\begin{equation}
\begin{tabular}{lcc}
 Model  & varying $m_e$  & $\Lambda$CDM  \\
\hline  \ensuremath{\Omega_{b}h^{2}}  &  0.0228808  &  0.0226626 \\
 \ensuremath{m_{e}/m_{e0}}  &  1.01705  &  1 \\
 \ensuremath{H_{0}}  &  71.2286  &  68.4235 \\
 \ensuremath{\Omega_m}  &  0.286295  &  0.301459 \\
 \ensuremath{Y_{P}^{{\rm {BBN}}}}  &  0.248037  &  0.246972\\
 \ensuremath{10^{5}D/H}  &  2.4326  &  2.46674 \\
\hline  
 \ensuremath{\chi_{\mathrm{Cooke}}^{2}}  &  3.89684  &  2.59375 \\
 \ensuremath{\chi_{\mathrm{Aver}}^{2}}  &  0.613477  &  0.267478 \\
 \ensuremath{\chi_{\mathrm{CMB}}^{2}}  &  2776.91  &  2777.93\\
 \ensuremath{\chi_{\mathrm{H0}}^{2}}  &  3.96685  &  21.9865 \\
 \ensuremath{\chi_{\mathrm{BAO}}^{2}}  &  9.95825  &  5.37674 \\
 \ensuremath{\chi_{\mathrm{prior}}^{2}}  &  3.49538  &  4.57984 \\
\hline  
 \ensuremath{\chi_{\mathrm{total}}^{2}}  &  2798.84  &  2812.73 
\end{tabular}
\end{equation}
\caption{ The best-fit $\chi^{2}$ of the varying $m_{e}$ model for other cosmological models 
and its comparison to the $\Lambda$CDM model. Here, we take 
$N_{\mathrm{eff}}^{\mathrm{SM}}=3.046$~\cite{Mangano:2005cc} of the default value in the \texttt{camb}. }
\label{Table:chi^2} 
\end{table}
We also show the prior dataset dependence in Table~\ref{Tab:68}.

\section{Summary}
\label{sec:summary}

We have investigated the BBN constraints on a time-variable electron mass model as a solution to the Hubble tension. 
There are two ways that the BBN prediction can be influenced. 
One is caused by the modification of the $p^+ \leftrightarrow n$ weak interaction rate by the dependence of the electron mass, which turns out not to be significant.
The other is due to the variation of $\Omega_b h^2$, because inferred cosmological parameters including $\Omega_b h^2$ are different from those in the $\Lambda$CDM to keep the CMB power spectrum unchanged for a shorter sound horizon scale at the recombination.
The inferred smaller $\Omega_b h^2$ suppresses $D/H$.
Nevertheless, the BAO constrains the cosmological parameter more significantly than the BBN does, 
as shown in Table~\ref{Table:chi^2} .

\begin{table}[htbp]
\begin{tabular}{lccc}
 Dataset  &  CMB+BAO & CMB+BAO+BBN & CMB+BAO+BBN+R21 
\\
\hline  \ensuremath{\Omega_{b}h^{2}}  &  \ensuremath{0.02250\pm 0.00017} & \ensuremath{0.02243\pm 0.00016} &  \ensuremath{0.02273 \pm 0.00014} \\
\ensuremath{m_{e}/m_{e0}} & \ensuremath{1.0048\pm 0.0065}& \ensuremath{1.0028\pm 0.0064} & \ensuremath{1.0182\pm 0.0048}\\
\ensuremath{H_{0}}  &  \ensuremath{68.5\pm 1.2}& \ensuremath{68.0 \pm 1.1}  &  \ensuremath{70.99 \pm 0.79}\\
\ensuremath{Y_{P}^{{\rm {BBN}}}}  &  \ensuremath{0.24719^{+0.00038}_{-0.00045} }& \ensuremath{0.24705^{+0.00035}_{-0.00043}} & \ensuremath{0.24804 \pm 0.00030}\\
\ensuremath{10^{5}D/H}  &  \ensuremath{2.498\pm 0.030} & \ensuremath{2.510 \pm 0.029} &  \ensuremath{2.458\pm 0.024}\\
\hline  
\end{tabular}
\caption{68\% limits for various datasets}
\label{Tab:68}
\end{table}

\section*{}
This work is supported in part by the Japan Society for the Promotion
of Science (JSPS) KAKENHI Grants No.~19K03860, No.~19K03865 and No.~21H00060 (O.S.)
 and JST SPRING, Grant No. JPMJSP2119 (Y.T).


\end{document}